# EXtensible Animator for Mobile Simulations: EXAMS

Livathinos S. Nikolaos, University of Patras

libathin@ceid.upatras.gr, nikos@livathinos.gr

*Abstract*— **One of the most widely used simulation environments for mobile wireless networks is the Network Simulator 2 (NS-2). However NS-2 stores its outcome in a text file, so there is a need for a visualization tool to animate the simulation of the wireless network. The purpose of this tool is to help the researcher examine in detail how the wireless protocol works both on a network and a node basis. It is clear that much of this information is protocol dependent and cannot be depicted properly by a general purpose animation process. Existing animation tools do not provide this level of information neither permit the specific protocol to control the animation at all. EXAMS is an NS-2 visualization tool for mobile simulations which makes possible the portrayal of NS-2's internal information like transmission properties and node's data structures. This is mainly possible due to EXAMS extensible architecture which separates the animation process into a general and a protocol specific part. The latter can be developed independently by the protocol designer and loaded on demand. These and other useful characteristics of the EXAMS tool can be an invaluable help for a researcher in order to investigate and debug a mobile networking protocol.**

*Index Terms*— **Visualization tools, network simulator 2 (NS-2), mobile networking simulations.**

## I. INTRODUCTION

In recent years there is an increase in mobile devices with networking capabilities. As a result there is a need for the research community to examine and design network protocols and applications for mobile networks. Interestingly, much of this work is done with the help of network simulation environments.

According to an earlier survey [5] one of the most widely used network simulator is the Network Simulator 2 (NS-2) [12] and this trend has been enhanced since then. In a typical NS-2 usage scenario the user describes the network configuration and parameters (like mobility and traffic patterns), feeds them in NS-2 input and after the simulation is finished NS-2 stores the simulation results in a text file being called the *trace file*. Usually trace files contain enormous amounts of data in text, encoded according to a given specification. Thus there is need for a tool that transforms the trace file to a visual playback of the simulation process. This tool could enable the researcher to use his visual system, an indisputable advantage to his effort in analyzing and debugging the network protocol. Animation may include the network layout, nodes' mobility patterns, packet transfers, details about data transmissions, values of nodes' internal data structures and so on. Moreover we want that tool to visualize not only a fixed set of protocols but have an extensible architecture, making it suitable to work with future protocol implementations.

At this paper we present a visualization tool for NS-2 mobile simulations, being called the *EXtensible Animator for Mobile Simulations* (EXAMS). EXAMS goal is to help the researcher understand the mobile routing protocol with an exact visual representation of the NS-2 simulation outcome, while on the same time enables him to control the animation process. What makes EXAMS different from other NS-2 animators is its extensible architecture which permits the protocol designer to specify the information shown during the simulation playback and how the program responds to user events. At technical level EXAMS can be seen as an independent visualization platform which is able to animate each separate protocol through *protocol extensions*. As each extension is developed by the protocol designer himself, EXAMS makes it possible to depict a more detailed level of information like transmission details and values of node's internal data structures. EXAMS specializes on the visualization of the routing and agent layer of NS-2 wireless simulations.

Our intension is to give freely EXAMS to the research community and it will be soon available via the World Wide Web as an open source project. The remaining part of this publication is organized as follows. Section II will describe other related work. Section III will give an overview of the EXAMS from the user's perspective. Section IV will examine the EXAM's internal architecture and discuss some design decisions. Section V will provide an extensive usage example and finally Section VI will close this paper.

## II. RELATED WORK

At this section we will describe in short other known tools for the visualization of NS-2's simulations results and justify why we believe that there is a real need for another tool.

Network AniMator (NAM) [2] is the official visualization tool for NS-2. However although NAM is good for wired networks it has not been extended to support wireless networks. Thus NAM is able to animate only nodes' movements and energy pulses when a node transmits a packet.

An early attempt to animate NS-2 wireless simulations was the *ad-hockey* [13] which was introduced by the CMU Monarch project [4] as an accompanying visualization tool for ns-2 mobile simulations. However this effort has been discontinued since 1999 and current versions of NS-2 do not include ad-hockey anymore.

A more recent work is the *interactive NS-2 protocol and environment configuration tool* (iNSpect) [6]. iNSpect fully supports wireless simulations and according to its developers there is an intention to support wired simulations at a forthcoming version. The main characteristic of iNSpect is its internal event builder which enables it to handle many types of input files and keep its own flow of event processing. More specifically, iNSpect accepts normal NS-2 trace files, mobility files used by NS-2 and a special type of trace file being called *vizTrace* file. This flexibility enables iNSpect to be used as a validation tool for mobility files bypassing the time consuming process of simulation. Additionally iNSpect can validate NS-2 itself as it is possible to check if NS-2 handles the mobility pattern the right way. Finally iNSpect can show network partitions and calculate traffic statistic reports.

However iNSpect provides very few details about the simulation process during the animation. It does not depict any data related with the current transmission, like the type of packet being processed, its size, ip source and destination, or other details provided by the routing protocol. It also does not have the ability to introspect the internal state of the nodes like the values of protocol specific data structures. Finally there is not any provision to let the specific protocol affect the animation process.

*Huginn* [10] is another environment for NS-2 wireless simulations. Its main characteristics are the usage of 3D graphics and a sophisticated scheme for the presentation of network related data. Using Huginn's graphical environment, the user can specify a data evaluation flowchart which defines the suitable filtering and aggregation of the initial data. This way it is able to make visual abstractions about the simulation process and enable the user to focus on issues he is interested in. However Huginn still suffers from the same limitations of the previous tools, as it is not able to introspect the nodes' internal data structures and does not allow the routing protocol to take control of the visualization process.

While the above visualization tools try to present the overall network (as EXAMS does), there are some other tools which concentrate on specific issues of the mobile environment. For instance [8] and [9] present *YAVISTA,* a graphical tool which compares network simulators (like NS-2 and GloMoSim [11]) about their implementation of the IEEE 802.11 MAC layer. This tool creates a static image representing the information exchange at the data link layer in the form of timeline. However YAVISTA does not accept normal trace files, but needs to have the data in an xml based neutral format. Thus significant modifications to the tracing module of each simulator are needed.

### III. OVERVIEW

EXAMS receives the NS-2 trace file as its main input. It can animate mobile nodes' movements, playback packets exchange at routing and agent level, give user the ability to freely move forward and backwards to simulation time, keep statistics about send/receive/forward/drop data at network and node basis and visualize network partitions and nodes' radius.

The goal is to help the user understand network's functionality both at a high level and on a node basis. At a high level user may be interested to see the overall layout and the movement patterns of the nodes, network partitions and radio range coverage as well as be informed about traffic statistics. On a node basis user may want to focus on a particular packet transmission and examine further details about that, or want to investigate the exact state of a specific node. We think that it is not possible to provide the user this level of information based on general assumptions about the underlying mechanisms of the network. It is obvious that each protocol has its own unique packet types, data structures, internal variables and states. For instance one protocol may use routing tables while another may not. Thus a proper visualization of the simulation process should be done according to protocol's *interpretation* and it is the protocol's developer responsibility to do it the right way.

In line with the above idea, EXAMS does not handle the visualization itself but provides the protocol's developer a *visualization infrastructure* which enables him to complete the animation process according to his protocol's needs. As a result EXAMS can be thought of as a basic infrastructure together with a set of protocols' implementations. When a specific simulation is to be visualized, EXAMS searches its protocols' implementations and chooses the right one according to the protocol that had been occupied by NS-2 during the simulation. Technically this is possible only due to EXAMS extensible architecture, where each routing protocol is an extension to the main program and can be developed and loaded a posteriori. We will further examine EXAMS architecture and programming model at the next section.

While other visualizers put much effort on the graphics, EXAMS first priority is the *exact* visualization of the simulation process. Additionally, there is no extra event builder or interpolation for the positions of the nodes between subsequent events, but each visualization event corresponds to exactly one line of the trace file in a one-to-one relationship. This choice makes EXAMS a good solution for the step by step examination and verification of existing protocols as well as the development and debugging of new ones.

Moreover EXAMS incorporates many user friendly features that will help the researcher to control the way the simulation looks and runs. Visualization's appearance can be fully controlled by the graphical user interface and it is

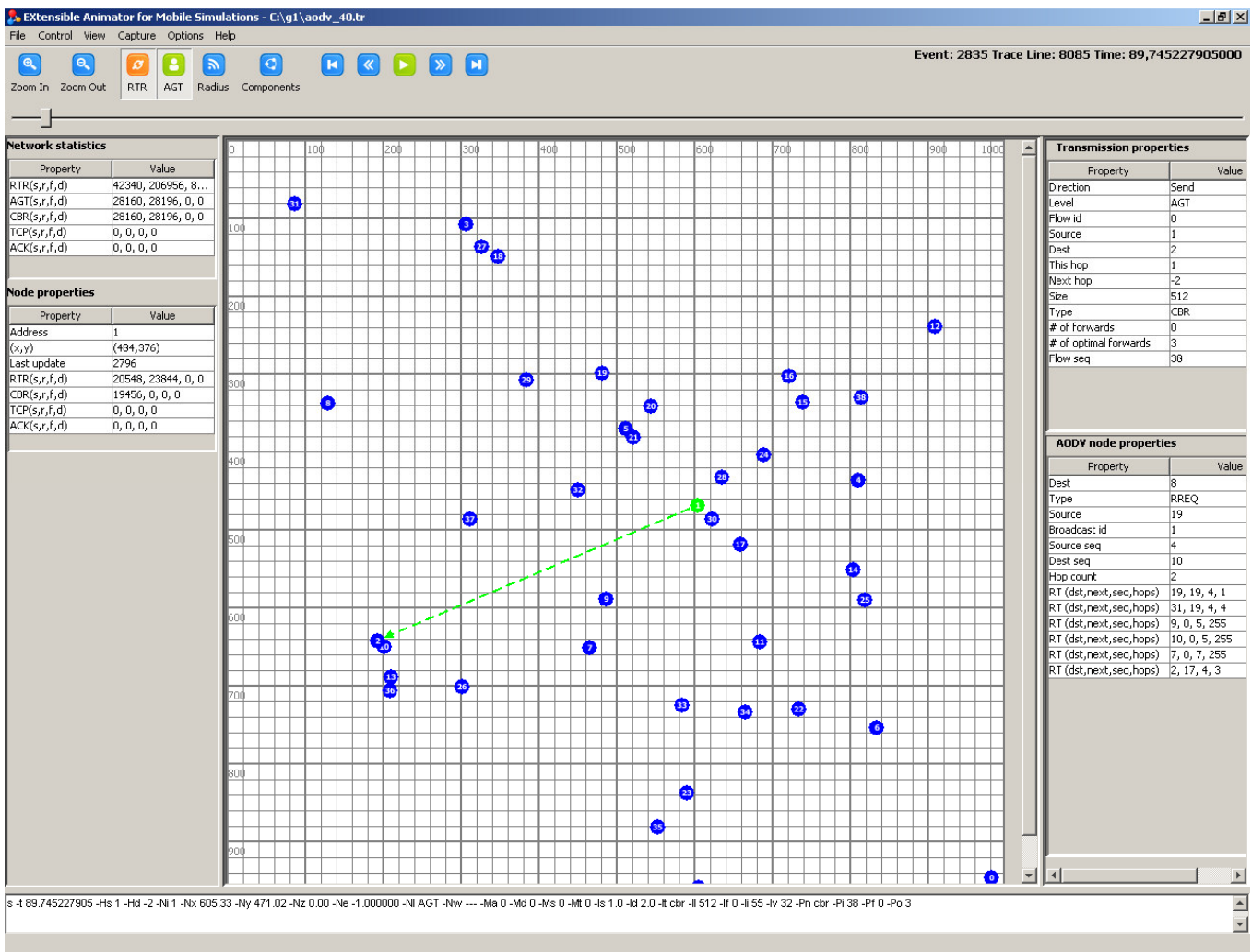

**Figure 1: EXAMS Environment**

possible to save or load the presentation's preferences to an external file. Other useful features are:

- Fast jump to specific line of the trace file.
- Zoom in /out capability.
- Filtering of network traffic according to routing and/or agent layer.
- Coloring of the network partitions according to given radio range.
- Screenshots capture to external image file.
- Color and shape codes for data transmission.
- Extensive usage of keyboard shortcuts.

As regards the color and shape codes, there is the convention to use fat line arrows for agent traffic and slim ones for routing traffic, as well as dashed lines for data send (or forward) and solid for data receive. The color of the lines/nodes is configurable and will change according to a send, receive, forward, drop or broadcasting event.

Figure 1 depicts a screenshot of the entire environment of the EXAMS tool. At the main region the network layout is shown, according to the color code described previously. The user can zoom in / out, get current position coordinates and take screenshots of network's current state into an external image file. The left side bar involve the network statistics and node properties tables which include statistics for the packets being transmitted on a network and node basis, as well as general information about the nodes like network address and position coordinates. On the right side bar there are the transmission properties and routing protocol properties tables. The former include information about a packet transmission like source and destination nodes, packet size, intermediate hops, network level and type of protocol involved. The latter concern issues related to the routing protocol occupied. Additionally the user can see in text the line of the trace file which corresponds to current event at the bottom of the window. The event slider on the top enables the user to make arbitrary event jumps.

All the information described above is automatically updated at each next event and the user is able to see the last known state of a random node by simply clicking on it. However the real power of EXAMS is its inherent ability to give the routing protocol designer the freedom to visualize protocol specific information in a simple and efficient way. This affects the information being shown at the right hand side bar (as shown in Figure 1).

The next section will provide more insight on the internal architecture of the EXAMS visualization tool, the adopted programming model and some interesting performance optimizations that have been taken into account at the design of the system.

## IV. EXAMS INTERNALS

### A. System prerequisites

EXAMS is a Java [3] desktop application (Java SE – version 1.6.0_02) and has been tested to work under the Linux (Fedora 9) and Windows (XP, Vista) operating systems. Normally it will work on every system that supports Java SE. We opt for Java as the programming environment for the EXAMS because, among others, it has the following advantages:

- It is a rich application development framework.
- It supports the building of Graphical User Interfaces (through Swing/AWT).
- The only prerequisite for Java SE applications is the Java Runtime Environment (JRE) which is normally very easy to be installed (in many cases it is already pre-installed). There are no dependencies with other libraries or need for extra settings.
- Java applications are easily deployed (through jar files).
- Cross platform support.
- Wide acceptance by the academic and research community.
- It is open source and free.

Thus given that the JRE has been properly installed, EXAMS can be deployed and used instantly without the need for any extra configuration.

### B. Architecture overview

The most outstanding feature of EXAMS is its ability to extend its functionality and incorporate new routing protocols. Figure 2 depicts a high level overview of EXAMS extensible architecture. We can see that EXAMS receives as input the NS-2 trace file, a proper implementation of the occupied routing protocol and the user's visualization preferences. The NS-2 trace file provides the simulation data which constitute the basis for the visualization, the routing protocol implementation encapsulates the semantics of the protocol which controls how the simulation data are presented and finally the visualization preferences are a simple way to handle additional animation issues.

NS-2 trace file is produced automatically by the NS-2 as a result of the simulation process, so it is a responsibility of the NS-2 protocol developer. As the trace file is the primary source for the visualization process, it is absolutely necessary to provide EXAMS all the necessary data for a proper presentation of the protocol under examination. Fortunately the NS-2 trace file layered architecture permits each protocol to add its own information at each related event, or in other words to add something at the corresponding line in the trace file. This enables the protocol designer to store a string representation of protocol's related data, like packet types, node's internal data structures, state information and so on. Then as the trace file is fed to EXAMS input, it is able to get back this information and visualize it. However this may be an issue for existing NS-2 implementations which export to the trace file only a small port of the internal protocol data. For instance in EXAMS a user may be interested in the records of the routing table inside a node, information which in general is not included in the trace file.

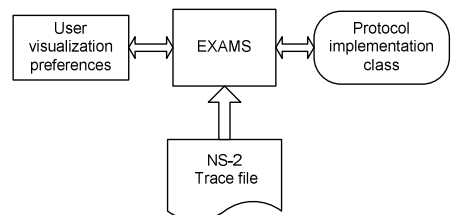

**Figure 2: EXAMS architecture overview**

As has been previously told, the EXAMS protocol implementations are extensions to the main infrastructure. Programmatically a protocol implementation for EXAMS is a Java class. As EXAMS parses the trace file it identifies the name of the routing protocol and expects to find a class file with exactly the same name under a specific classpath. This Java class is the protocol's implementation for EXAMS and is used to handle everything that is protocol dependent. This procedure is possible due to Java Reflection Mechanism [15], which enables a Java application to load and use a class at the run time given the class file location. In simple the goals of an EXAMS protocol implementation is the proper parsing of the protocol related data inside the trace file and their visualization according to protocol's semantics. Next sections will describe in detail how to develop a protocol implementation for EXAMS and use its infrastructure to visualize it.

Finally the user can set his visualization preferences as additional information to the animation process. These include node and transmission line colors, terrain dimensions, filters, directory options and other minor issues. User can set his preferences inside the graphical environment of EXAMS and has the ability to store and load his preferences to an external XML file.

### C. Detailed description of EXAMS architecture

Figure 3 illustrates a more detailed view of EXAMS architecture and the way it uses the protocol implementation to get the desired results.

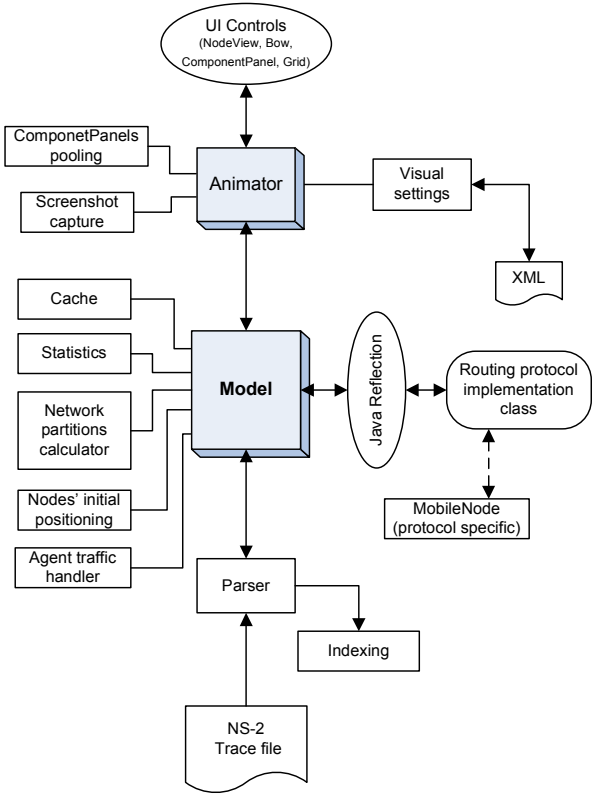

**Figure 3: EXAMS architecture**

As it seems at the core of EXAMS lays the Model class which provides the principal functionality of event traversing. We want the user to be able to see the state of the whole network at every event of the simulation and make random event jumps. On the other hand an NS-2 trace file is only a set of lines, each one corresponding to a simulation event which in turn is related to one node only (e.g packet sender, or receiver). In order to visualize the whole network at every event, Model has an internal array to store the state of the nodes of the network. As a next line of the trace file is processed, the Model can update the state of the corresponding node but keep the remaining nodes at the previous known state. However this model implies the limitation that events should be processed sequentially as the next event is an update to network's current state. At a next section we will see some optimizations to overcome these limitations and enable the user to move freely to a future or a past event and start the animation process from that point and on. The Model is also responsible for a number of things like calling Parser to read the trace file, load the Protocol implementation class, keep statistics of the network traffic, handle agent packets, calculate network partitions and other.

Another important class is the Animator class which is responsible for the visualization process itself. It uses the Model to know each time the state of each node and creates a visual representation of the network. This includes the nodes themselves, unicast and multicast packet transmission at routing and agent level and the outlining of network partitions with different colors. Animator can plot either on the screen or create a screenshot of the network into an image file.

One issue has to do with positions of the nodes at the start of the simulation, because there are some routing protocols which do not specify the initial nodes layout. EXAMS handles this problem with an early positioning of the nodes. It scans the trace file and sets the initial position of each node as indicated by the first event related to that node. Early positioned nodes are shown grayed until their initial settlement.

As regards statistics EXAMS keeps packet counters at both network and node level. Each node has two sets of counters; one set is used for the routing layer and the other one for the agent layer. At each layer EXAMS counts the number of bytes being sent, received, forwarded or dropped. A same couple of counter sets is used to measure the total number of bytes being transferred into the network at routing and agent layer. Both these sets of counters are available to the protocol developer and can be used as the basis for the calculation of more complex statistics.

*D. Protocol development programming model*

From a programming point of view every protocol implementation for EXAMS should implement the Protocol Java interface or alternatively extend the DummyProtocol class which is a naive implementation of the same interface.

Table 1 shows the methods of the Protocol interface. The most important methods are createNode, updateNode, copyNode and notifyEvent. One major responsibility of a class implementing the Protocol interface class is to create the MobileNode objects through the createNode method. Each MobileNode object represents a mobile node and contains all the data about the state of the node. Moreover a protocol developer can subclass the MobileNode class and create a more suitable version for the specific protocol with its unique data structures and functionality. After the initial creation of the MobileNode objects, EXAMS calls periodically the updateNode method in order to update node's stored information and keep up to date the network. CopyNode method is a necessity for the system in order to make deep[1] copies of the MobileNode objects. At last the notifyEvent method is called by the Animator class each time the user sends a UI event like a mouse click on a node or a transmission arrow. This in turn enables the Protocol class to further handle the user event according to protocol's semantics.

| Return value | Method |
|---|---|
| String | getName() |

---

[1] According to [16] a deep copy is "a copy of a data structure duplicating not only the structure itself, but all structures to which is linked".

| | |
|---:|:---|
| MobileNode | `createNode(int address, double x, double y, HashMap<String, String> properties)` |
| MobileNode | `updateNode(MobileNode n, HashMap<String, String> properties, int eventIndex)` |
| MobileNode | `copyNode(MobileNode n)` |
| void | `notifyEvent(VisualEvent e)` |

**Table 1: Protocol Java Interface**

The advantage of this architecture is that it permits the protocol designer to handle everything that is protocol specific. He can define the nodes, the way they are created, what should happen when a related simulation event appears and what is the response to a user event. However it is also important to make the whole process as simple as possible for the protocol designer and hide the complexity of EXAMS internals. Thus although Figure 3 depicts the internal architecture of EXAMS it is not a necessary knowledge for the routing protocol designer. Figure 4 shows the protocol development programming model under EXAMS.

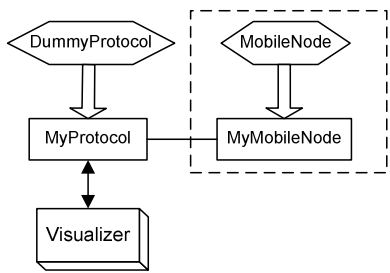

**Figure 4: Protocol programming model**

According to Figure 4 there is a routing protocol named MyProtocol, so a Java class with the same name should be implemented. This class can either implement directly the Protocol interface (as described in Table 1) or extend the DummyProtocol. Optionally protocol's designer can implement a subclass of MobileNode to encapsulate protocol's internal data structures as described above. The most interesting part of Figure 4 is the Visualizer class. It is a wrapping of EXAMS engine and aims to relieve protocol developer from the complexity of EXAMS internals. A protocol class can obtain a Visualizer object by calling a static method and use it as a service provider. Visualizer provides access to main screen's tables, simulation events, packet counters, visual settings and in general whatever a protocol may need to handle the services provided by EXAMS infrastructure.

*E. Performance optimizations*

As stated at previous sections, an NS-2 trace file is a set of simulation events. Each event is associated with one network node and every event is an update to the previous network state. This way a transition to a random event requires the sequential application of every event starting from the initial network state up to the desired event. Of course this is an inefficient technique especially when the user wants to replay a specific region of the simulation or make random event jumps. EXAMS overcomes this inherent inefficiency of NS-2 trace files by occupying a network state caching scheme. The idea is to keep a deep copy of the nodes' configuration in order to load it immediately at a future user jump. We call this deep copy as a network *snapshot*. Moreover snapshots can be used as intermediate starting points for a sequential update. When the user wants to jump to a certain event that is not present in the cache, EXAMS searches its cache to find the closest previous snapshot and starts a sequential update from this event and on.

However we still have to decide about the caching scheme itself; which snapshots should be kept and how they should be organized. We need an efficient mechanism being able to eliminate cache misses while on the same time reduce the memory overhead caused by the storage of the deep copies. Thus it is important to make good predictions about the user's next jumps. Interestingly enough, we can apply the *locality principle* on predicting the user's behavior. According to [1] "The locality principle flows from human cognitive and coordinative behavior. The mind focuses on a small part of the sensory field and can work most quickly on the objects of its attention. People gather the most useful objects close around them to minimize the time and work of using them. These behaviors are transferred into the computational systems we design". This means that the user will tend to jump to event locations that are local to the current one either temporal or spatial. Temporal locality refers to events that the user had previously visited and spatial locality to events that are close to the current one. In cases affected by the locality principle it has been shown ([1]) that the best practice is to keep a local cache with a least recently used (LRU) replacement algorithm. This way cache keeps always objects with small distance (temporal or spatial) and removes those ones with big distance.

According to the previous analysis EXAMS correlates every simulation event to an array of MobileNode describing the state of all the nodes. Then each MobileNode array is packed with the corresponding event id and is stored to the local cache. Thus when the user asks for a specific event, EXAMS can respond immediately and load the suitable MobileNode array given that this event id is present in cache, otherwise a sequential update should be applied.

Additionally EXAMS keeps a local cache not only for the simulation events but for the associated network partitions too. We want to give to the user a visual hint for the network's overall radio range coverage. As nodes are moving around they separate the network to network components, or partitions. EXAMS calculates the union of the radio range of the nodes that belong to the same partition and visualizes it as the partition's radio range. Each partition is shown with a different color. It is interesting to notice that NS-2, in general, uses a more realistic model for radio propagation than that of a circular and universally equal radio range. However as the trace file does not include any information about nodes' radio range, EXAMS bases its calculations on the approximated circular range given by the user. Figure 7 illustrates the partitions of a network in EXAMS.

Another important optimization has to do with the Parser itself. NS-2 trace files can be very large and thus make it impossible to be entirely loaded from disk to memory. On the other hand we need reasonable response times for the file access process. EXAMS makes possible to satisfy these constraints by the occupation of two different indexes. The first index correlates each line of the trace file with its offset from the beginning of the file and the second one maps each EXAMS event with the corresponding line number. Thus when a specific event needs to be loaded, EXAMS initiates a two step process. First it uses the second index to find the corresponding line of the trace file and then uses the first index to find the starting and ending byte of that line inside the file stream. This way we can fetch quickly a trace line without having to load the entire file into memory. Moreover an internal buffer eliminates the need for file accesses in cases of subsequent calls for the same event.

## V. EXAMS PARADIGM

At this section an extensive paradigm of EXAMS usage is to be shown. This paradigm will be based on the well known routing protocol for mobile ad hoc networks AODV [7], as it is implemented in the NS-2 simulator. More specific we want to visualize the following points:

1. General network data (network event types, network level, packet's size, nodes' positions, ip source and destination, current hop etc).
2. AODV data provided in standard NS-2 trace files (see [14] for the exact trace data format).
3. AODV internal state data, not provided in standard NS-2 trace files.

The first two points have the advantage that the data are included in standard NS-2 trace files, so the only thing should be done is a proper parsing of the trace file. As regards the third point this needs modifications at the NS-2 code in order to export internal AODV data structures to the trace file. For the sake of simplicity we populate only the routing table of each AODV node. Each record of the routing table is exported as the tuple <destination address, sequence number, number of hops, next hop for the destination>. Table 2 depicts an example of the part of the trace file that is related to the routing protocol, according to the "newtrace" file format (see [14]). At the end of the modified trace there is the "Prt" directive which carries information about the entries of the routing table for the corresponding node.

As has been shown in section D we should encapsulate all the functionality for the visualization of the AODV in a Java class being named with the same name (AODV.class). This class should implement the Protocol interface, or equivalently subclass the DummyProtocol class. We opt for the second one, and override the methods createNode, copyNode, updateNode, getName and notifyEvent. The role of these methods has been explained in section D. Moreover it would be convenient to subclass the MobileNode class to provide a customized class for AODV nodes. This class will include the internal AODV data being exported in NS-2 trace files (the routing table in case of our example) as well as the parser for the AODV part of NS-2 trace file.

| Original part of NS-2 trace file | -P aodv -Pt 0x2 -Ph 1 -Pb 1 -Pd 8 -Pds 0 -Ps 7 -Pss 4 -Pc REQUEST |
| --- | --- |
| Part of NS-2 trace file after modifications | -P aodv -Pt 0x2 -Ph 1 -Pb 1 -Pd 8 -Pds 0 -Ps 7 -Pss 4 -Pc REQUEST -Prt (8,0,255,0)(1,5,255,0) |

**Table 2: Comparison of NS-2 trace file before and after AODV extensions for internal data introspection**

Figure 5 depicts an NS-2 simulation for the AODV routing protocol of 40 nodes at 1000*1000m terrain. The figure shows node 11 receiving a routing packet from node 6.

Figure 6 is a screenshot from the various tables with data which are available to the EXAMS user at the event depicted in Figure 5. Figure 6 is separated into four parts and each one represents four tables with data about the current event (see Figure 1 for the exact layout of EXAMS tables). In part 1 we can see the network statistics. It includes the total number of bytes sent in the routing and agent layer. In case of the agent layer EXAMS further analyzes it either to CBR or to TCP/ACK packets. In part 2 we can see some general data about the current node. As seen it includes the node's address, its current location and its internal counters for the layers described before. The "Last update" field refers to the last event concerning the current node, so it is an indication of the freshness of the data about that node. In part 3 the properties of the current transmission are shown. This include the direction and the network level of the packet, its flow id, ip source and destination, current and next hop and finally packet size and type. Eventually in part 4 the routing protocol specific data are shown. As we can see, it involves the type of the packet according to the AODV semantics, the source and destination address, the source and destination sequence number and the hop counter. Up to this point all these data are provided by the standard trace files. However after the modification we have made to AODV tracing, it is possible to see the routing table of the current node. It is shown in the last lines of part 4. Each record is a tuple according to Table 2. For instance we can see that at current event, node 11 has a known route for node 19 which is 2 hops away and the next hop should be node 1.

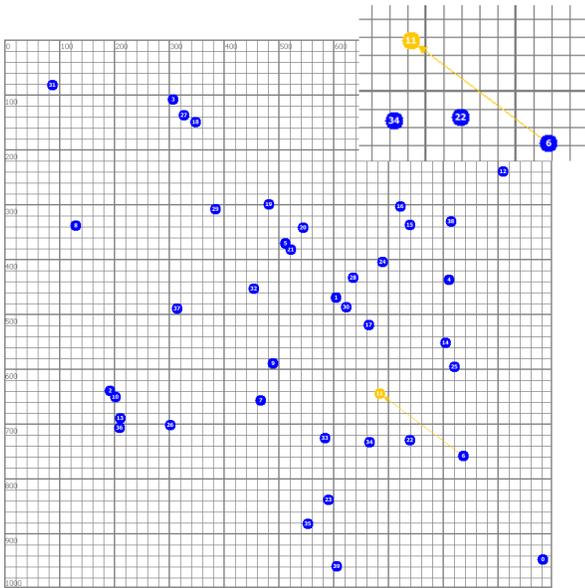

**Figure 5: EXAMS terrain screenshot of an AODV network
(the inlaid image is a close up for the area in interest)**

Another characteristic of the network is the partitions it is split. Figure 7 illustrates each partition of the network with a different color which covers the whole area of the partition. As nodes are moving around, the partitions are recalculated and the user is able to see the whole radio range coverage.

Summing up, we saw how we can use EXAMS to study the mobile ad-hoc protocol AODV. As AODV had already been implemented in NS-2, we added some extra code to its tracing module in order to include current node's routing table at the trace file. At a newly developed protocol this step could be part of the initial implementation and thus don't comprise an extra burden. Then we implemented the AODV Java class which handles the AODV specific issues inside EXAMS and the AODVNode class which incarnates an AODV node. It is interesting to mention that the Java code needed for the protocol implementation classes is less than 300 lines for both the two classes. Figures 5, 6 and 7 provide some idea of the plenty of data provided about the simulation process as well as the visual environment of EXAMS. We believe that EXAMS is a useful tool which can exploit the human visual system and help the researcher to understand how a wireless protocol behaves under the NS-2 simulator while on the same time provides him all the vital information to conclude about the protocol internals. This information is crucial when each step of the simulation process should be investigated in detail as in case of debugging an under development protocol or making a rigorous study of an existing one.

## VI. CONCLUSION

With the rapid growth of network enabled mobile devices there is an increased need for wireless applications. NS-2 is a widely used tool for mobile network simulations and EXAMS can reveal the simulation process in visual. EXAMS is a helpful tool especially for the user who wants to know in detail the response of the protocol while on the same time be able to see protocol's internal variables and data structures. EXAMS extensible architecture makes it possible to apply this level of visualization to any routing protocol while its simple programming model minimizes the needed programming effort.

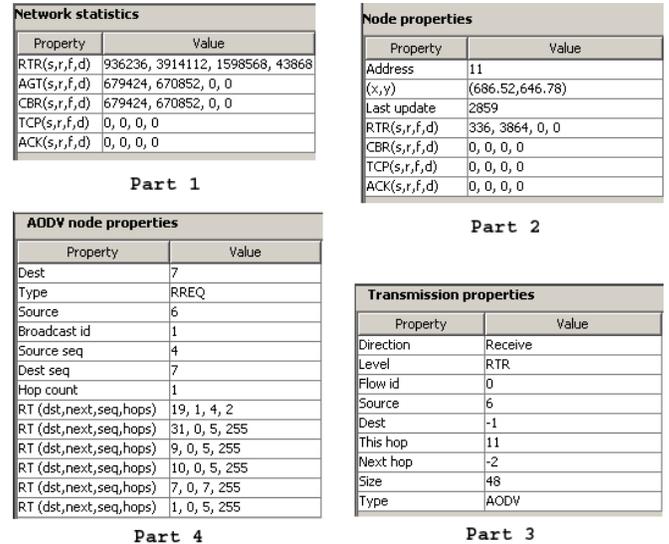

**Figure 6: Simulation properties**

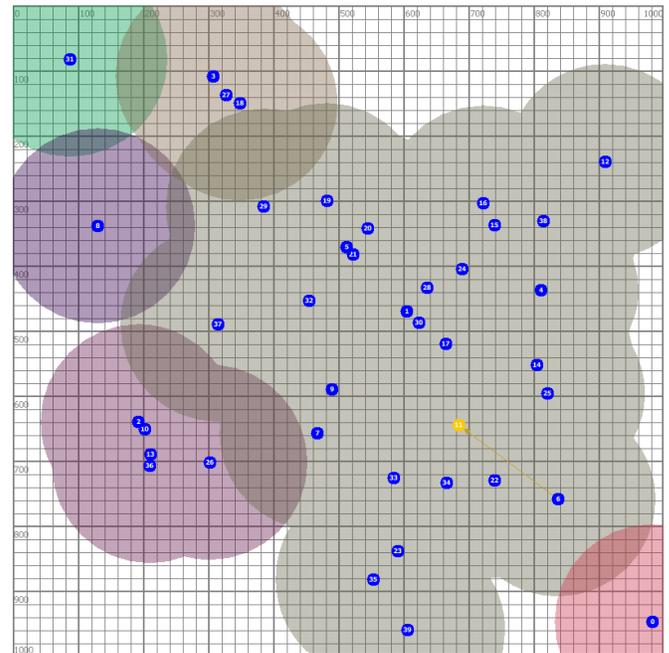

**Figure 7: Network partitions**

In a common usage scenario the following steps enable the user to fully exploit EXAMS capabilities:
- Export to NS-2 trace file all the information which is to be seen during the visualization like nodes' internal data structures.
- Develop Java code which handles the visualization details about the protocol specific issues.
- Load the trace file.

Our intension is to give freely EXAMS to the research community and it will be soon available via the World Wide Web as an open source project.

## VII. ACKNOWLEDGMENT

Livathinos S. Nikolaos would like to thank …